\documentclass[11pt,a4paper,english,nofootinbib,,superscriptaddress]{revtex4}
\usepackage{lmodern}

\usepackage[T1]{fontenc}
\usepackage[latin9]{inputenc}
\usepackage{babel}
\usepackage{color}
\usepackage{amsmath}
\usepackage{graphicx}
\usepackage{amssymb}
\usepackage{esint}
\usepackage[unicode=true, pdfusetitle,
 bookmarks=true,bookmarksnumbered=false,bookmarksopen=false,
 breaklinks=false,pdfborder={0 0 1},backref=false,colorlinks=false]
 {hyperref}
\usepackage{amsmath}
\usepackage{amssymb}
\def\b{\begin{equation}} \def\e{\end{equation}}
\def\bd{\begin{displaystyle}} \def\ed{\end{displaystyle}}
\def\ba{\begin{array}} \def\ea{\end{array}}

\def\bee{\begin{enumerate}}
\def\eee{\end{enumerate}}

\def\1{\mbox{I\hspace{-.15em}1}}

\def\Z{\mbox{Z\hspace{-.3em}Z}}

\def\R{{\rm I\hspace{-.15em}R}}

\def\b{\begin{equation}}
\def\e{\end{equation}}
\def\bee{\begin{enumerate}}
\def\eee{\end{enumerate}}

\makeatletter
\usepackage{latexsym}\usepackage{bm}

\makeatother
\begin{document}

\title{Quantum de Sitter-black hole in ambient space formalism}

\author{M.V. Takook}
\email{takook@razi.ac.ir} 
\affiliation{Department of Physics, Razi University,
Kermanshah, Iran}
\affiliation{Department of Physics,
Science and Research branch, \\ Islamic Azad University, Tehran,
Iran}

\date{\today}

\begin{abstract}

Two important problems in studying the quantum black hole, namely the construction of the Hilbert space and the definition of the time evolution operator on such Hilbert space, are discussed using the de Sitter background field method for an observer far from the black hole. This is achieved through the ambient space formalism.  Remarkably, in this approximation (distant observer), the theory preserves unitarity and analyticity, it is free from any infrared divergence, and it renders a quantum black hole entropy that turns out to be finite.

\end{abstract}

\maketitle
\vspace{0.5cm}
{\it Proposed PACS numbers}: 04.62.+v, 03.70+k, 11.10.Cd, 98.80.H
\vspace{1.5cm}

\section{Introduction}

In a previous paper the microscopic origin of the de Sitter (dS) entropy in the context of quantum field theory (QFT) in ambient space formalism was considered and the total number of quantum states was calculated  \cite{ta13}. For this purpose the Hilbert spaces for QFT in dS ambient spaces were constructed.  In fact the QFT in ambient space formalism enables us to consider QFT in a rigorous mathematical framework based on the analyticity of the complexified pseudo-Riemannian manifold and the group representation theory (see  \cite{ta14} for a review). With this formalism the author has been able to obtain some interesting results \cite{morrota,parota,ta14,rota05,paenta,gata17}.

In the present paper, building upon the work on the Hilbert spaces and the time evolution operator obtained previously, we show that this formalism can also be used to solve the well-known problems of unitarity in the quantum dS-black hole and the microscopic origin of the dS-black hole entropy.  

The organisation of this paper is as follows: In section II we shall introduce the notation and two independent Casimir operators of the dS group.  Section III is devoted to a brief review of the one-particle Hilbert space; we find the total number of quantum states, which, remarkably, turns out to be finite.  The time evolution operator is introduced in Section IV.  The quantum dS-black hole, in particular its quantum state and entropy, are discussed in section V.  Conclusions and an overview of this work, with possible extensions, are presented in the final section.

\section{NOTATION and TERMINOLOGY} 
\label{notations}

The dS space-time can be identified with a 4-dimensional hyperboloid embedded in the 5-dimensional Minkowski space-time:
     \b \label{dSs} M_H=\{x \in \R^5| \; \; x \cdot x=\eta_{\alpha\beta} x^\alpha
 x^\beta =-H^{-2}\},\;\; \alpha,\beta=0,1,2,3,4, \e
where $\eta_{\alpha\beta}=$diag$(1,-1,-1,-1,-1)$ and $H$ is the Hubble parameter. The dS metric is \b \label{dsmet}  ds^2=\eta_{\alpha\beta}dx^{\alpha}dx^{\beta}|_{x^2=-H^{-2}}=
g_{\mu\nu}^{dS}dX^{\mu}dX^{\nu},\;\; \mu=0,1,2,3,\e 
where $X^\mu$ is the set of 4-space-time intrinsic coordinates on the dS hyperboloid. In this paper we use the 5-dimensional Minkowski space-time  $x^\alpha$ with the condition ($x\cdot x=-H^{-2}$) which together constitute the ambient space formalism.  For simplicity, we shall hereafter set  $H=1$ wherever it is convenient and immaterial. 
The dS group is defined by: 
 $$ SO(1,4)=\left\lbrace \Lambda \in GL(5,\R)| \;\; \det \Lambda=1,\;\; \Lambda \eta \Lambda^t= \eta \right\rbrace,$$ where $\Lambda^t$ is the transpose of $\Lambda$. The action of the dS group on the intrinsic coordinates $X^\mu$ is a non-linear operation, however on the ambient space coordinate $x^\alpha$, it is linear:
$$ x'^\alpha=\Lambda^\alpha_{\;\;\beta} x^\beta, \;\;\;\Lambda \in SO(1,4) \Longrightarrow x\cdot x=x'\cdot x'=-H^{-2}. $$  The ambient space coordinate $x^\alpha$ can be defined by a $4\times 4$ matrix ${\bf X}$:
\b \label{mx44} {\bf X}=\left( \begin{array}{clcr} x^0I & \;\;{\bf x} \\ {\bf x}^\dag & x^0I \\    \end{array} \right),\e
$I$ is the $2\times 2$ unit matrix where
\b \label{quat} {\bf x}=\left( \begin{array}{clcr} x^4+ix^3 &\;\; ix^1-x^2 \\ ix^1+x^2 & \;\;x^4-ix^3 \\    \end{array} \right).\e 
Note that  ${\bf x}$ can be represented by a quaternion ${\bf x}\equiv(x^4, \vec x)$ with the norm: 
$$|{\bf x}|=({\bf x}{\bf  \tilde{ x}})^{\frac{1}{2}}=\sqrt{(x^1)^2+(x^2)^2+(x^3)^2+(x^4)^2},$$
where ${\bf \tilde{ x}}\equiv(x^4,-\vec x)$ is the quaternion conjugate of ${\bf x}$ \cite{tak}. In the matrix notation, we have ${\bf  \tilde{ x}}={\bf x}^\dag$ and the norm can be written as $|{\bf x}|^2=\frac{1}{2} \mbox{Tr}( {\bf x}{\bf  x}^\dag) $. The matrix ${\bf X}$ may be expressed in an alternative form which is  more convenient for our considerations in this paper: 
\b \label{gamax} \not x =\eta_{\alpha\beta}\gamma^\alpha x^\beta={\bf X}\gamma^0=\left( \begin{array}{clcr} x^0I & \;\;-{\bf x} \\ {\bf  \tilde{ x}} & -x^0I \\    \end{array} \right),\;\;\; \not x\not x =x\cdot x \;\1,\;\;\frac{1}{4} \mbox{Tr}\left(\not x\not x\right) =x \cdot x  ,\e 
where $\1$ is a $4\times 4$ unit matrix and the five matrices $\gamma^{\alpha}$ satisfy the following conditions \cite{bagamota}:
$$\gamma^{\alpha}\gamma^{\beta}+\gamma^{\beta}\gamma^{\alpha}
=2\eta^{\alpha\beta}\qquad
\gamma^{\alpha\dagger}=\gamma^{0}\gamma^{\alpha}\gamma^{0}.$$ 
We also use the following representation for $\gamma$ matrices in this paper \cite{bagamota}:
$$ \gamma^0=\left( \begin{array}{clcr} I & \;\;0 \\ 0 &-I \\ \end{array} \right)
      ,\;\;\;\gamma^4=\left( \begin{array}{clcr} 0 & I \\ -I &0 \\ \end{array} \right) , $$ \b \label{gammam}
   \gamma^1=\left( \begin{array}{clcr} 0 & i\sigma^1 \\ i\sigma^1 &0 \\
    \end{array} \right)
   ,\;\;\gamma^2=\left( \begin{array}{clcr} 0 & -i\sigma^2 \\ -i\sigma^2 &0 \\
      \end{array} \right)
   , \;\;\gamma^3=\left( \begin{array}{clcr} 0 & i\sigma^3 \\ i\sigma^3 &0 \\
      \end{array} \right),\e
where $\sigma^i$ $(i=1,2,3)$ are the Pauli matrices. In this representation the matrix $\not x$ transforms by the group $Sp(2,2)$ according to:
$$ \not x'=g\not x g^{-1}, \;\;\;g \in Sp(2,2),\;\;\mbox{Tr} \left( \not x\not x\right)=\mbox{Tr} \left( \not x'\not x'\right) \Longrightarrow x\cdot x=x'\cdot x'=-H^{-2}.$$
The group $Sp(2,2)$ is: \b \label{sp22} Sp(2,2)=\left\lbrace g=\left( \begin{array}{clcr} {\bf a} & {\bf b} \\ {\bf c} & {\bf d} \\    \end{array} \right), \;\; \det g=1 , \;\; \gamma^0 \tilde{ g}^t \gamma^0=g^{-1} \right\rbrace . \e 
The elements ${\bf a},{\bf b},{\bf c}$ and ${\bf d}$ are quaternions and $Sp(2,2)$ is the universal covering group of $SO(1,4)$ \cite{bagamota}:
 \b SO_0(1,4)  \approx Sp(2,2)/ \Z_2,\;\; \;\;\Lambda_{\alpha}^{\;\;\beta}\gamma^\alpha=g \gamma^\beta g^{-1} .\e 
For the $4 \times 4$ matrix $g$, we have $\tilde{ g}^t=g^\dag$. 
 
In this paper, two different types of homogeneous spaces are used to construct the UIR of the dS group: the quaternion ${\bf u}\equiv(u^4, \vec u)$ with norm $|{\bf u}|=1$ which is called three-sphere $S^3$ (or ${\bf u}$-space), and the quaternion ${\bf q}\equiv(q^4, \vec q)$ with norm $|{\bf q}|<1$ which is called "closed unit ball" or for simplicity "the unit ball $B$" (or ${\bf q}$-space). 

These two homogeneous spaces can be regarded as the sub-spaces of the positive cone $C^+$, which is defined as $C^+=\left\lbrace \xi \in \R^5|\;\; \xi\cdot \xi=0,\;\; \xi^{0}>0 \right\rbrace$. Thus the null 5-vector $\xi^\alpha=(\xi^0, \vec \xi, \xi^4)\in C^+$ may be written, uniquely, as:
\b \label{2orbit} \xi^\alpha_u \equiv (\xi^0, \xi^0 \; {\bf u}),\;|{\bf u}|=1;\;\;\; \xi^\alpha_B \equiv (\xi^0 , \xi^0\; \coth \kappa \; {\bf q}),\; \; |{\bf q}|=|\tanh \kappa|=r<1.\e 
Since $\xi\cdot \xi=0$, from the mathematical point of view,  $\xi^0$ is completely arbitrary, {\it i.e.} $\xi^\alpha$ is scale invariant.

If we choose ${\bf q}=r{\bf u}$ with $r<1$, we obtain $$ \xi^\alpha_B = (\xi^0 , \xi^0\; \coth \kappa \; {\bf q})\equiv (\xi^0, \xi^0 \; {\bf u})=\xi^\alpha_{u}.$$

In this notation, $x\cdot\xi$ may be written in the following form:
\b \label{x.xi} \not x \not \xi + \not \xi \not x=2\; x\cdot \xi\; \1 \Longrightarrow x\cdot\xi= \frac{1}{4} \mbox{Tr}\not x \not \xi \,,\e 
whence both are invariant under the action of the dS group:
\b \label{x.xit}  x'\cdot \xi'  = x\cdot \xi,\;\;  \mbox{Tr} \not x' \not \xi'= \mbox{Tr}\not x \not \xi.\e Here the $\xi$-space plays the role of the energy-momentum $k^\mu$ in the Minkowski space-time and it can be chosen for massive field as \cite{brgamo}:
$$ \xi^\alpha_u \equiv \xi^0\left(1, \frac{\vec{k}}{k^0} , \; \frac{H \nu}{k^0}\right) ,\;\; k^0 \neq 0,\; \nu \neq 0,\;\;\; (k^0)^2- \vec{k}\cdot \vec{k}=(H\nu)^2, $$ 
where $\nu$ is the principal series parameter which will be further explained in the next section. For the massless field, $\xi$, we have:
$$ \xi^\alpha_B = \xi^0 \left( 1, \; \frac{{\bf q}}{r}\right)\equiv \xi^0\left(1,  \frac{\vec{k}}{k^0} , \; \frac{H}{k^0}\right) ,\;\; k^0 \neq 0,\; \;\;\; (k^0)^2- \vec{k}\cdot \vec{k}=H^2.$$
In the null curvature limit, we obtain precisely the massive and the massless energy momentum,  $(k^0)^2- \vec{k}\cdot \vec{k}=m^2$ and $(k^0)^2- \vec{k}\cdot\vec{k}=0$ respectively.

The dS group has two Casimir operators: a second-order Casimir operator,
 \b \label{casi1} Q^{(1)}=-\frac{1}{2}L_{\alpha\beta}L^{\alpha\beta},\;\;  \alpha, \beta=0,1,2,3,4, \e and a fourth-order Casimir operator,
      \b Q^{(2)}=-W_\alpha W^\alpha\;\;,\;\;W_\alpha =\frac{1}{8}
      \epsilon_{\alpha\beta\gamma\delta\eta} L^{\beta\gamma}L^{\delta\eta},\e
where  $\epsilon_{\alpha\beta\gamma\delta\eta}$ is the familiar anti-symmetric tensor in $\R^5$ and $L_{\alpha\beta}$ are the infinitesimal generators of the dS group,  $L_{\alpha\beta}=M_{\alpha\beta}+S_{\alpha\beta}$. In the ambient space formalism the orbital part, $M_{\alpha\beta}$, is
          \b \label{genm} M_{\alpha \beta}=-i(x_\alpha \partial_\beta-x_\beta
      \partial_\alpha)=-i(x_\alpha\partial^\top_\beta-x_\beta
        \partial^\top_\alpha),\e
where $\partial^\top_\beta=\theta_\beta^{\;\;\alpha}\partial_\alpha$ and $\theta
_{\alpha\beta}=\eta_{\alpha\beta}+H^2x_\alpha x_\beta$ is the projection tensor on the dS hyperboloid. Integer spin fields can be represented by the symmetric tensor fields of rank $l$,
 ${\cal K}_{\gamma_1...\gamma_l}(x)$ and the spinorial action reads
\cite{gaha}:
       \b \label{gens} S_{\alpha \beta}^{(l)}{\cal K}_{\gamma_1......\gamma_l}=-i\sum^l_{i=1}
          \left(\eta_{\alpha\gamma_i}
        {\cal K}_{\gamma_1....(\gamma_i\rightarrow\beta).... \gamma_l}-\eta_{\beta\gamma_i}
          {\cal K}_{\gamma_1....(\gamma_i\rightarrow \alpha).... \gamma_l}\right),\e
where $(\gamma_i\rightarrow\beta)$ means that $\gamma_i$ are to be replaced
by $\beta$.

For the $j=2$ integer, the Casimir operator $ Q^{(1)}_2$ acts on a rank-$2$ symmetric tensor filed ${\cal K}_{\alpha_1 \alpha_2 }(x)\equiv {\cal K}(x) $ as \cite{gaha,taazba}:
 \b \label{casimirl} Q_2^{(1)}{\cal K}(x)=Q_0^{(1)}{\cal K}(x)-2\Sigma_1 \partial x \cdot {\cal K}(x)+2\Sigma_1 x \partial\cdot
           {\cal K}(x)+2 \eta {\cal K}'(x)-6{\cal K}(x),\e
where
\b Q_0^{(1)}=-\frac{1}{2}M_{\alpha \beta}M^{\alpha \beta}=-H^{-2}\partial^\top\cdot\partial^\top \equiv -H^{-2} \square_H\;.\e  $\square_H$ is the Laplace-Beltrami operator of the dS space-time. ${\cal K}'$
is the trace of the rank-$2$ tensor field ${\cal K}(x)$; and $\Sigma_p$ is the non-normalized symmetrization operator:
         \b {\cal K}'=\eta^{\alpha_{1}\alpha_2}
          {\cal K}_{\alpha_1\alpha_{2} },\e
         \b \label{sigmap} (\Sigma_p AB)_{\alpha_1...\alpha_l}=\sum_{i_1<i_2<...<i_p}
          A_{\alpha_{i_1}\alpha_{i_2}...\alpha_{i_p}}
          B_{\alpha_1...\not\alpha_{i_1}...\not\alpha_{i_2} ...\not\alpha_{i_p}...\alpha_l}\;.\e
The tensor field ${\cal K}(x)$ on the dS hyperboloid is transverse ($ x \cdot {\cal K}=0$) and is a homogeneous function of the variables $x^{\alpha}$ with degree $\lambda$ \cite{dirds}: $$ x\cdot\partial {\cal K}(x)=\lambda {\cal K}(x), \;\; \mbox{or}\;\; {\cal K}(lx)=l^\lambda{\cal K}(x).$$
The field equations may thus be written \cite{gagata}: \b \label{fexs} \left( Q^{(1)}_2- \left<Q^{(1)}_{2,p}\right>\right){\cal K}(x)=0\, ,\e
where $\left<Q^{(1)}_{2,p}\right>$ is the eigenvalue of the Casimir operator which will be defined presently.

\section{Unitary irreducible representation and Hilbert space}

The Casimir operators commute with the generators of the dS group and as a consequence, they are constant in each unitary irreducible representation (UIR) of the dS group. The UIRs of the dS group are classified by the eigenvalues of the Casimir operators \cite{tho,new,dix,tak}:
             \b \label{qe1} Q^{(1)}_{j,p}=\left(-j(j+1)-(p+1)(p-2)\right)I_d\equiv Q^{(1)}_{j} , \e
           \b Q^{(2)}_{j,p}=(-j(j+1)p(p-1))I_d, \e
$I_d$ is the identity operator. Three types of representations, corresponding to the different values of the parameters $j$ and $p$, exist \cite{ta14}. Here, only the spin-$2$ field is considered:
\begin{itemize}
\item{Principal series representation},
 \b\label{ps}  j=2,\;\;\;p=\frac{1}{2}+i \nu,\; \nu \in \R\, , \; \nu \geq 0\, ;
         \;\;\left<Q^{(1)}_{2}\right>=-\frac{15}{4}+\nu^2\, .\e
\item{Discrete series representation},
\b    j=2\, ,\;\;\; p=1,2\, ;
         \;\; \left<Q^{(1)}_{2}\right>=-6\, , \;\;\; -4. \e
\item{Complementary series representation},
          \b  j= 2\, ,\;\;\;\;\;0<p-p^2=\mu^2 <\frac{1}{4}\, ;
         \;\;  \left<Q^{(1)}_{2}\right>=-4+\mu^2\,.\e
\end{itemize}

The principal series representation of the dS group was constructed in the compact homogeneous three-sphere space, $S^3$ or ${\bf u}$-space \cite{tak}:
\b \label{principals} U^{(j,p)}(g)\left|{\bf u},m_j;j,p \right\rangle=|{\bf c}{\bf u}+{\bf d}|^{-2(1+p)}\sum_{m_j'}D^{(j)}_{m_jm_j'}\left(\frac{({\bf c}{\bf u}+{\bf d})^{-1}}{|{\bf c}{\bf u}+{\bf d}|} \right) \left|g^{-1}\cdot {\bf u},m_j';j, p \right\rangle,  \e
where $g^{-1}\cdot {\bf u}= ({\bf a}{\bf u}+{\bf b})({\bf c}{\bf u}+{\bf d})^{-1}$ with $g^{-1}=\left( \begin{array}{clcr} {\bf a} & {\bf b} \\ {\bf c} & {\bf d} \\    \end{array} \right) \in Sp(2,2)$. $D_{m_jm_j'}^{(j)}$ furnish a certain representation of the $SU(2)$ group in a $(2j+1)-$dimensional Hilbert space $V^j$:
$$ D^{(j)}_{mm'}({\bf u})=\left[\frac{(j+m)!(j-m)!}{(j+m')!(j-m')!}\right]^{\frac{1}{2}} \times $$
\b \label{su2} \sum_n \frac{(j+m')!}{n!(j+m'-n)!}\frac{(j-m')!}{(j+m-n)!(n-m-m')!}  u_{11}^{n}u_{12}^{j+m-n}u_{21}^{j+m'-n}u_{22}^{n-m-m'} .\e
The representation $U^{(j,p)}$ act on an infinite dimensional Hilbert space ${\cal H}^{(j,p)}_{u}$:
$$ \left|{\bf u},m_j;j,p \right\rangle \in {\cal H}^{(j,p)}_{u}, \;\;\; \xi_u=(\xi^0, \xi^0 {\bf u}),\; \;\; \xi^0 >0,\;\;\;|{\bf u}|=1, \;\;\; -j\leq m_j\leq j.$$ 

The UIR of the dS group for discrete series is constructed on the unit ball homogeneous space $B$ or ${\bf q}$-space \cite{tak}:
$$ T^{(j_1,j_2,p)}(g)\left|{\bf q},m_{j_1},m_{j_2};j_1,j_2,p\right\rangle=|{\bf c}{\bf q}+{\bf d}|^{-2(1+p)} \times $$ \b \label{dseris2}\sum_{m_{j_1}'m_{j_2}'}D^{(j_1)}_{m_{j_1}'m_{j_1}}\left( \frac{({\bf a}+{\bf b}\bar {\bf q})^{-1}}{|{\bf c}{\bf q}+{\bf d}| }\right)
D^{(j_2)}_{m_{j_2}m_{j_2}'}\left(\frac{({\bf c}{\bf q}+{\bf d})^{-1}}{|{\bf c}{\bf q}+{\bf d}|} \right)
 \left|g^{-1}\cdot {\bf q},m_{j_1}',m_{j_2}';j_1,j_2,p\right\rangle.  \e
In this case, one has an infinite dimensional Hilbert space  ${\cal H}^{(j_1,j_2,p)}_{q}$,
$$ \left|{\bf q},m_{j_1},m_{j_2};j_1,j_2,p\right\rangle \in {\cal H}^{(j_1,j_2,p)}_{q},\;\;\; {\bf q}=r{\bf u} \in \R^4, |{\bf q}|=r<1, \;\; |{\bf u}|=1,\;\; -j\leq m_j\leq j.$$
The discrete series representations $T^{(j,0,p)}$ and $T^{(0,j,p)}$ are proportional to the two representations $\Pi^+_{j,p}$ and $\Pi^-_{j,p}$ in the Dixmier notation \cite{dix}:
\b \label{dis0jp} T^{(0,j;p)}(g)\left|{\bf q},m_j;j,p\right\rangle=|{\bf c}{\bf q}+{\bf d}|^{-2p-2}\sum_{m_j'}D^{(j)}_{m_jm_j'}\left(\frac{({\bf c}{\bf q}+{\bf d})^{-1}}{|{\bf c}{\bf q}+{\bf d}|} \right)\left|g^{-1}\cdot {\bf q},m_j';j,p\right\rangle, \e and 
\b \label{dis0jp2} T^{(j,0;p)}(g)\left|{\bf q},m_j;j,p\right\rangle=|{\bf c}{\bf q}+{\bf d}|^{-2p-2}\sum_{m_j'}D^{(j)}_{m_j'  m_j}\left(\frac{({\bf a}+{\bf b}\bar {\bf q})^{-1}}{|{\bf c}{\bf q}+{\bf d}|} \right)\left|g^{-1}\cdot 
{\bf q},m_j';j,p\right\rangle. \e
The representations $\Pi^\pm_{j,j}$ in the null curvature limit correspond to the Poincar\'e massless field with helicity $\pm j$.

As regards (\ref{dis0jp}) and (\ref{dis0jp2}), one can show that the two representations with the values $p_1$ and  $p_2$ are unitary equivalents ($p_1+p_2=1$) \cite{tak}:
\b \label{ued} {\cal S} \left|{\bf q},m_j;j,p \right\rangle=\left|{\bf q},m_j;j,1-p \right\rangle,\;\;\;\; {\cal S}T^{(0,j;p)}(g) {\cal S}^\dag=T^{(0,j;1-p)}(g), \e
with ${\cal S}{\cal S}^\dag=1$. It implies that the representation $T^{(0,j;p)}(g)$ for the $p=\frac{1}{2},1,\frac{3}{2},2,...$ values is the unitary equivalent of $T^{(0,j;1-p)}(g)$ with $p=\frac{1}{2},0,-\frac{1}{2},-1,...$. The eigenvalues of the Casimir operators for these two set of value or under the transformation $p \longrightarrow 1-p$, do not change.

The complementary series representations can only be associated with the tensor fields with $j=0,1,2,..$. These are constructed on ${\bf u}-$space or ${\bf q}-$space \cite{dix,tak}. Among these representations, only the scalar representation with $j=p=0$ has corresponding Poincar\'e group representation in the null curvature limit. This representation relates to the massless conformally coupled scalar field \cite{brmo}. The massless conformally coupled scalar field is the building block of the massless fields in dS space, since all the other massless spin fields can be constructed using this field. The complementary series representation with $j=p=0$ is defined as \cite{tak}:
\b \label{comserp} U^{(0,0)}(g)\left|{\bf u},0;0,0 \right\rangle=|{\bf c}{\bf u}+{\bf d}|^{-2} \left|g^{-1}\cdot {\bf u},0;0, 0\right\rangle. \e
The scalar products in these Hilbert spaces are presented by a function ${\cal W}({\bf u},{\bf u}')$ which is defined on $S^3\times S^3$ \cite{tak}. The equivalent  unitary representation (\ref{comserp}) is corresponds to $j=0,\; p=1$, this is:
\b \label{comserpueq} U^{(0,1)}(g)\left|{\bf u},0;0,1 \right\rangle=|{\bf c}{\bf u}+{\bf d}|^{-4} \left|g^{-1}\cdot {\bf u},0;0, 1\right\rangle. \e 

One of the most important result that follow from this formalism is that the total volume of the homogeneous spaces, in which the UIR of the dS group are constructed, namely ${\bf q}$-space or ${\bf u}$-space ($\xi$-space), is finite:
$$ \int_{S^3}d\mu({\bf u})=1, \;\; \int_{B}d\mu({\bf q})=\frac{\pi^2}{2},$$
where $d\mu({\bf u})$ is the Haar measure or $SO(4)$-invariant normalized volume on the three-sphere $S^3$ and  $d\mu({\bf q})=2\pi^2r^3drd\mu({\bf u})$ is the Euclidean measure on the unit ball $B$ \cite{tak}. We have used the definition  ${\bf q}=r{\bf u}\in B$ with ${\bf u}\in S^3, \;r<1$.

Here the two $x$- and $\xi$-spaces play a role similar to space-time and energy-momentum in Minkowskian space-time. The existence of a maximum allowed length for an observable implies, by virtue of the Heisenberg uncertainty principle, the existence of a  minimum size in the $\xi$-space (or the parameters in Hilbert space). Each point in $\xi$-space represents a vector in Hilbert space and, mathematically,  the number of points is infinite. From the finiteness of the total volume of $\xi$-space and the existence of a minimum length in $\xi$-space (by virtue of the Heisenberg uncertainty principle), we deduced that the total number of points is physically finite. Thus, although one can mathematically define an infinite dimensional Hilbert space, nonetheless a minimum length in $\xi$-space makes the total number of quantum states physically finite. For the principal series (${\cal H}^{(j,p)}_{u}$), we have \cite{ta13}:
\b {\cal N}_{{\cal H}^{(j,p)}_{u}} =(2j+1) \int_{S^3} d\mu(\xi_u)=f(H,j,\nu,\xi^0)=\mbox{finite value},\e
and for discrete series ${\cal H}^{(j_1,j_2,p)}_{q}$, it is
\b {\cal N}_{{\cal H}^{(j_1,j_2,p)}_{q}}=(2j_1+1)(2j_2+1) \int_B d\mu(\xi_B)=\mbox{finite value}.\e
The total number of quantum states is a function of $H$, $p$, $j$ and $\xi^0$. This result is due to the existence of a minimum length and the compactness of the homogeneous spaces in which the Hilbert spaces (or the UIR) are constructed. Thus the entropy for this Hilbert spaces is finite \cite{ta13}.

\setcounter{equation}{0} 
\section{The time evolution operator} 
\label{mmcfield1}

In order to define the time evolution operator $U(t)$, such that $\vert \alpha , t\rangle=U(t) \vert \alpha \rangle$, we must first define  the time of an observer or the coordinate system. We chose the static coordinate system with:
 \b \left\{\begin{array}{clcr} x^0&=\sqrt{H^{-2}-r^2}\sinh Ht_s \\                    
                      x^1&=\sqrt{H^{-2}-r^2}\cosh Ht_s \\
                       x^2&=r\cos \theta  \\
                                  x^3&=r \sin \theta\cos\phi \\   
                                   x^4&=r\sin\theta\sin \phi\\            
         \end{array} \right.\e
where $-\infty<t_s<\infty\;,\;0\leq r<H^{-1}\;,\;0\leq \theta\leq\pi\;,\;0\leq \phi< 2\pi$. This coordinate system does not cover all the de Sitter hyperboloids. In this coordinate system the metric is:        
         \b ds^2=\left(1-r^2H^2\right)dt_s^2-\left(1-r^2H^2\right)^{-1}dr^2- r^2 \left(d \theta^2 +\sin^2
              \theta d\phi^2\right)\, ,\e 
and the time-like Killing vector field is:
$$   B_0=i\frac{r\cosh Ht_s}{\sqrt{H^{-2}-r^2}}\partial_{t_s}.$$
              
In ambient space formalism, the time translation operator in the two-dimensional ($\theta=\phi=0$) de Sitter space-time is:
      \b g(H\tau )=\left(\begin{array}{clcr} \cosh H\tau  & \sinh H\tau & 0 \\  
                                                \sinh H\tau & \cosh H\tau & 0 \\
                                                 0 & 0 & 1 \\
                          \end{array} \right), \;\;\; \;\;  g(H\tau)x(t_s,r)=x(t_s+\tau,r).\e
The time translation is the subgroup $SO(1,1)$ whose generator can be obtained as follow:
\b  g(H\tau)\equiv \left(\begin{array}{clcr} \Lambda (H\tau)  &   0 \\  
                                               0 & 1 \\
                          \end{array} \right)\,   ,\;\;\; \Lambda (H\tau)  \in SO(1,1)\, ,\e
                          \b \Longrightarrow  J=iH^{-1}\left.\frac{d}{d\tau} \Lambda(H\tau )\right\vert_{\tau=0}=i\left(\begin{array}{clcr}  0 & 1 \\  
                                                1 & 0  \\
                          \end{array} \right)\Longrightarrow \Lambda(H\tau) =e^{-iH\tau J}.\e
                          
{\bf Exercise:} Fined the representation of $J$ in the de Sitter Hilbert space  $\left|{\bf u},m_j;2,p \right\rangle$ ($\label{ues}\langle{\bf u}',m'_j;2,p\vert J \left|{\bf u},m_j;2,p \right\rangle=?$). Also calculate $ U(\tau)\left|{\bf u},m_j;2,p \right\rangle=?$.
Verify that  $UU^\dag=1$ and $U(\tau_1)U(\tau_2)=U(\tau_1+\tau_2)$.

The relevant exercise for the case of a scalar field was considered in \cite{brmo}. Thus we have a time evolution operator which is unitary.

\setcounter{equation}{0}
\section{Quantum black hole}

In this section, we use the background field method to study a black hole in the de Sitter space-time. The de Sitter metric in the statics coordinate system is:  
\b \left.ds^2\right\vert_{\mbox{dS}}=\left(1-H^2r^2\right)dt^2-\left(1-H^2r^2 \right)^{-1}dr^2- r^2 \left(d \theta^2 +\sin^2 \theta d\phi^2\right).\e 
And the de Sitter-Schwarzschild black hole metric is:
\b \left.ds^2\right\vert_{\mbox{dSB}}=\left(1-\frac{2GM}{r}-H^2r^2\right)dt^2-\left(1-\frac{2GM}{r}-H^2r^2 \right)^{-1}dr^2- r^2 \left(d \theta^2 +\sin^2 \theta d\phi^2\right).\e 
For an observer far from the black hole where the condition $Hr>\frac{GM}{r}$ holds, the metric can be divided into two parts: the de Sitter metric as the background, and a perturbation field $h_{\mu\nu}$:
\b g_{\mu\nu}^{\mbox{dSB}}=g_{\mu\nu}^{\mbox{dS}}+h_{\mu\nu}.\e
Then one can simply express the perturbation field $h_{\mu\nu}$, imposed on dS background, as: 
\b h_{\mu\nu}=g_{\mu\nu}^{\mbox{dSB}}-g_{\mu\nu}^{\mbox{dS}}.\e
In this approximation, the problem of the quantum black hole is reduced to that of quantizing the perturbation field $h_{\mu\nu}$ and defining the Hilbert space. This field is a rank-$2$ symmetric tensor, which was completely considered in the ambient space formalism in the previous articles \cite{taro15,ta09,gagata,tatafa,tata,petata,enrota}. Here we simply quote the important results of those articles which are necessary for calculating the entropy of the black hole from the Hilbert space, for an observer far from the black hole.

The metrics within the two formalisms are related according to:
\b \label{dsmet}  \left. ds^2\right\vert_{\mbox{dS}}=\eta_{\alpha\beta}dx^{\alpha}dx^{\beta}|_{x^2=-H^{-2}}=
g_{\mu\nu}^{dS}dX^{\mu}dX^{\nu},\;\; \mu=0,1,2,3,\e
where $X^{\mu}$ is the static coordinate system. The rank-$2$ symmetric tensor field $ {\cal K}_{\alpha\beta}$ in ambient space can be easily calculated from the following relation:
\b dX^\mu=\frac{\partial X^\mu}{\partial x^\alpha}dx^\alpha \Longrightarrow  {\cal K}_{\alpha\beta}=\frac{\partial X^\mu}{\partial x^\alpha}\frac{\partial X^\nu}{\partial x^\beta}h_{\mu\nu}.\e
 
 {\bf Exercise:} Use the condition $\vert h \vert ^2\approx 0$ ({\it i.e.} the linear approximation) to calculate: a) the rank-$2$ symmetric tensor $ {\cal K}_{\alpha\beta}$, b) the field equation for  $h_{\mu\nu}$, and c) the field equation for ${\cal K}_{\alpha\beta}$ (see \cite{tata}).
 
 From the field equation for ${\cal K}_{\alpha\beta}$, we conclude that there exist four possibilities:
 \begin{enumerate}
  \item  ${\cal K}$ may be a spin-$2$ massless elementary field which can be associated with the discrete series representation; the relevant field equation being \cite{ta09,taro15}:
$$ \left( Q_2^{(1)}+6\right){\cal K}_{\alpha\beta}=0,\;\;\;\mbox{or}\;\;\; Q^{(1)}_0{\cal K}_{\alpha\beta}=0.$$
In the discrete series representation there exists another spin-$2$ auxiliary field, which their field equation is \cite{petata}:
$$ \left( Q_2^{(1)}+4\right){\cal K}_{\alpha\beta}=0\, .$$
In these cases the Hilbert space is:
$$ \left|{\bf q},m_{j_1},m_{j_2};j_1,j_2,p\right\rangle \in {\cal H}^{(j_1,j_2,p)}_{q},\;\;\; {\bf q}=r{\bf u} \in \R^4, |{\bf q}|=r<1, \;\; |{\bf u}|=1,\;\; -2\leq m_j\leq 2.$$
We have two helicity: $j_1=2,\;j_2=0$ and $j_1=0,\;j_2=2$. ${\bf q}$ is a quaternion with the norm $\vert{\bf q}\vert <1$.
  \item   ${\cal K}$ may be a spin-$2$ massive elementary field which can be associated with the principal series representation. The field equation is \cite{ta14}:
$$ \left[ Q_2^{(1)}-\left(\nu^2-\frac{15}{4}\right)\right]{\cal K}_{\alpha\beta}(x)=0\, . $$
The Hilbert space is:
$$ \left|{\bf u},m_j;2,p \right\rangle \in {\cal H}^{(j,p)}_{u}, \;\;\; \xi_u=(\xi^0, \xi^0 {\bf u}),\; \;\; \xi^0 >0,\;\;\;|{\bf u}|=1, \;\;\; -2\leq m_j\leq 2.$$
   \item   ${\cal K}$ may be a spin-$2$ elementary field which can be associated with the complementary series representation. The field equation is \cite{ta14}:
$$ \left( Q_2^{(1)}+4+p^2-p\right){\cal K}_{\alpha\beta}=0\,.$$
The complementary series is constructed in the unit ball homogeneous space $B$ or ${\bf q}$-space \cite{tak}.

  \item   ${\cal K}$ may be a spin-$2$ field which is not an elementary field: rather, it is a composite field. In This case the Hilbert space is the tensorial product of different Hilbert spaces for each component.
\end{enumerate}

Thus the quantum state of a de Sitter black hole in the linear approximation in a general case may be formally written as:
$$\vert QBHS \rangle \propto \sum_{j,p,m_j}\int_{B}d\mu({\bf q})\cdots \sum_{j',p',m'_j}\int_{B}d\mu({\bf q'})  \, \times $$ \b  C({\bf q},j,p,m_j;\cdots;{\bf q'},j',p',m'_j) \, \left|{\bf q},m_j;j,p\right\rangle \otimes \cdots \otimes \left|{\bf q'},m'_j;j',p'\right\rangle\, , \e  
which is written thus for discrete series. Finding the explicit form of the coefficient $C$ is left to the reader as an exercise. In the case of principal and complementary series, the summation over $p$ is replaced with an integral.

Since the total volume of ${\bf q}$- or ${\bf u}$-space is finite,
$$ \int_{S^3}d\mu({\bf u})=1, \;\; \int_{B}d\mu({\bf q})=\frac{\pi^2}{2},\;\;\; d\mu({\bf q})=2\pi^2r^3drd\mu({\bf u}), $$
 and since a minimum length in these spaces exists (by virtue of the uncertainty principle), the total number of points in the ``one-particle'' Hilbert space becomes ``physically'' finite . It follows that the total number of quantum states in these Hilbert spaces is finite \cite{ta13}. For the de Sitter black hole due to the appearance of interaction, which is a bounded force classically, the de Sitter black hole entropy is:
\b S_{dSB}<S_{dS}=k_B \ln {\cal N}_{ren}.\e

Using equation (\ref{ues}), we obtain, for an observer far from the black hole, the time evolution operator of the black hole:
\b \vert QBHS , t \rangle=U(t)\vert QBHS \rangle\, ,\e
which {\it is} unitary. The representation of the time translation generator $J$ in the black hole Hilbert space is quite cumbersome. Obtaining the explicit form of the generator $J$ is left to the serious reader as an exercise.

\section{Conclusion and outlook}
\label{conclu}

The de Sitter ambient space formalism has permitted us to solve numerous problems of QFT in curved space time and quantum gravity \cite{ta14}. In this article we gave an outline of how the problems of the quantum state and its time evolution can be solved using the ambient space formalism for the quantum dS-black hole in the linear approximation. Interestingly, the entropy turns out to be finite too. Some calculations, which however do not change the results presented here, are left for the reader as exercises.

\vspace{0.5cm} 

{\bf{Acknowledgements}}:  The author wishes to express his particular thanks to J.P. Gazeau and S. Tehrani-Nasab for helpful discussions.

\end{document}